\documentclass[12pt]{article}
\pdfoutput=1
\usepackage{cite}
\usepackage{amssymb}
\usepackage{amsmath}
\usepackage{bm} 
\usepackage{graphicx}
\usepackage{color}
\usepackage{xcolor}
\usepackage{dsfont}
\usepackage{enumerate}
\usepackage{hyperref}
\usepackage{setspace}
\usepackage[latin1]{inputenc}
\usepackage{amsmath,amssymb,amsthm}

\usepackage{mathrsfs}
\usepackage{subfigure}
\usepackage{graphics}
\definecolor{nicered}{rgb}{0.7,0.1,0.1}
\definecolor{nicegreen}{rgb}{0.1,0.5,0.1}
\hypersetup{colorlinks,citecolor= nicegreen,linkcolor= nicered}

\newcommand{\be}  {\begin{equation}}
\newcommand{\ee}  {\end{equation}}

\def\e6{E(6)}
\def\10{SO(10)}
\def\21{SA(2) $\otimes$ U(1) }
\def\321{$\mathrm{SU(3) \otimes SU(2) \otimes U(1)}$ }

\def\422{SA(4) $\otimes$ SA(2) $\otimes$ SA(2)}
\def\roughly#1{\mathrel{\raise.3ex\hbox{$#1$\kern-.75em
      \lower1ex\hbox{$\sim$}}}} \def\lsim{\roughly<}
\def\gsim{\roughly>}

\def\lsim{\raise0.3ex\hbox{$\;<$\kern-0.75em\raise-1.1ex\hbox{$\sim\;$}}}
\def\gsim{\raise0.3ex\hbox{$\;>$\kern-0.75em\raise-1.1ex\hbox{$\sim\;$}}}

\nopagebreak
\allowdisplaybreaks
\begin{document}
\begin{titlepage}


  \newcommand{\AddrIPM}{{\sl \small School of physics, Institute for
      Research in Fundamental Sciences (IPM),\\ \sl \small
      P.O. Box 19395-5531, Tehran, Iran}}

  \newcommand{\AddrPSU}{{\sl \small Department of Physics; Department of Astronomy \& Astrophysics; Center for Particle and
Gravitational Astrophysics, The Pennsylvania State University, PA 16802, USA.\\ \sl \small}}

  \vspace*{0.5cm}
\begin{center}
{\bf{\large Lepton Flavor Violating Non-Standard Interactions \\ via Light Mediators }}\\ \vspace{.5cm}
  Yasaman Farzan$^a$~\footnote{e-mail address:{\tt yasaman@theory.ipm.ac.ir}} and Ian M. Shoemaker$^b$ \footnote{e-mail address:{\tt shoemaker@psu.edu}}
  \vspace*{0.4cm}\\
  $^a$\AddrIPM.
    \vspace*{0.4cm}\\
  $^b$\AddrPSU.
\end{center}
\vspace*{0.2cm}
\begin{abstract}
 \onehalfspacing

Non-Standard neutral current Interactions (NSIs) of neutrinos with matter can alter the pattern of neutrino oscillation due to the coherent forward scattering of neutrinos on the medium. This effect makes long-baseline neutrino experiments such as NO$\nu$A and DUNE a sensitive probe of beyond standard model (BSM) physics.  We construct light mediator models that can give rise to both lepton flavor conserving as well as Lepton Flavor Violating (LFV) neutral current NSI. We outline the present phenomenological viability of these models and future prospects to test them. We predict a lower bound on Br$(H\to \mu \tau)$ in terms of the parameters that can be measured by  DUNE and NO$\nu$A, and  show that the hint for $H\to \mu \tau$ in current LHC data can be accommodated in our model.
{A large part of the parameter space of the model is already constrained by the bound on Br$(\tau \to Z^\prime \mu)$ and by the bounds on rare meson decays and can be  in principle fully tested by improving these bounds. }
\end{abstract}
\end{titlepage}
\setcounter{footnote}{0}
\section{Introduction
}\label{sec:intro}

All observations and experimental evidence so far show that neutrinos only have weak interactions and can be well described within the Standard Model (SM) of particle
physics. It is however intriguing to ask  whether these rather mysterious particles can have any new interactions that have not been  so far detected. In particular, if there is a new neutral current interaction with matter fields ({\it i.e.,} with the electron or first generation quarks), it can affect neutrino propagation in matter~\cite{Wolfenstein:1977ue}. Thus physics beyond the Standard Model can be probed via neutrino oscillations in matter.  The Non-Standard neutral current Interaction (NSI) of neutrinos can be effectively described by the following four-Fermi interaction: \be \label{NSI}
\mathcal{L}_{NSI}=-2\sqrt{2}G_F\epsilon_{\alpha \beta}^{f P}
(\bar{\nu}_\alpha \gamma^\mu L\nu_\beta)(\bar{f}\gamma_\mu P ~f) \ee
where $f$ is the matter field ($u, \ d$ or $e$), $P$ is the
chirality projection matrix and $\epsilon_{\alpha \beta}^{f P}$ is a
dimensionless matrix describing the deviation from the SM.
 Surprisingly the upper bounds on $\epsilon_{\alpha \beta}$ from neutrino oscillation effects are rather weak
\cite{Tommy,deGouvea:2015ndi}.
The 90 \% C.L. current bounds from neutrino oscillation observations are \cite{Maltoni}:
\be \label{current-bounds}
|\epsilon_{e \mu}^{u(d)}|<0.09, \ \ \ \ \ |\epsilon_{e \tau}^{u(d)}|<0.14, \ \ \ \ \ |\epsilon_{\mu \tau}^{u(d)}|<0.01,
 \ee
and
$$  |\epsilon_{\mu \mu}^{u(d)}-\epsilon_{e e}^{u(d)} |<0.51 \ \ \ \ |\epsilon_{\mu \mu}^{u(d)}-\epsilon_{\tau \tau}^{u(d)} |<0.03  .$$
{Notice that these are model-independent constraints using only neutrino oscillation data. More model-dependent bounds arise from Tevatron and LHC data which can be stronger than those in Eq.~(\ref{current-bounds}) for mediator masses heavier than $\mathcal{O}(100~{\rm GeV})$ for some flavor combinations~\cite{Friedland:2011za,Franzosi:2015wha}. 

 In the near-term future, long baseline experiments such as NO$\nu$A  and the upcoming state-of-the-art DUNE experiment~\cite{DUNE,Acciarri:2015uup} will bring about unprecedented opportunities to probe these couplings further. Indeed, the recent establishment of the DUNE collaboration has created renewed interest in NSI~
\cite{Friedland:2012tq,deGouvea:2015ndi,Coloma:2015kiu,Masud:2015xva}.  Furthermore, improvements on NSI limits can be made by better measurements of the electron neutrino survival probability  near the MSW ``upturn'' around a few MeV~\cite{Friedland:2004pp,Palazzo:2011vg}. In fact, dark matter direct detection experiments may be able to offer improved determination of the behavior of solar electron neutrinos, and hence NSI~\cite{Billard:2014yka}.

The underlying UV-complete model giving rise to Eq.~(\ref{NSI}) has to respect the electroweak symmetry so it will in general also give rise to sizeable
new interactions of charged leptons \cite{Antush}, which tend to be much more strongly constrained \cite{Julian-Heeck}.
Further constraints on the underlying model are imposed by non-detection of the new mediator particle \cite{Julian-Heeck}, which we denote as $X$, whose exchange gives rise to new effective interactions \cite{Julian-Heeck}. Let us denote the coupling and mass of this new particle by $g_X$ and $m_X$. The $\epsilon$ parameters describing the deviation from the SM can be estimated as $\epsilon \sim (g_X^2/M_X^2) G_F^{-1}$ so to obtain $\epsilon\sim 1$, the ratio $g_X/m_X$ should be fixed to $\sim G_F^{1/2}$. Non-detection of the new particle can be explained in two limits: (i) $m_X \gg m_Z$; (ii) $g_X\ll 1$. In the former case, which has been the focus of most model builders, perturbativity of $g_X$ ({\it i.e.,} $g_X \leq 1$) implies that $\epsilon \ll 1$. That is we would not expect any sizeable non-standard effect on neutrino propagation.
Ref \cite{Me} suggests to invoke the second option with $g_X\sim 5 \times 10^{-5}$ and $m_X\sim 10$~MeV to obtain $\epsilon \sim 1$. Notice that since for neutrino propagation only forward scattering of matter is relevant ({\it i.e.,} $t$-channel diagrams in which $X$ is exchanged with $t=0$), we can use the effective coupling in Eq. (\ref{NSI}) even for neutrino energies much higher than $m_X$.  Ref. \cite{Me} introduces a consistent model for the so-called LMA-Dark solution  \cite{Dark} with   $\epsilon_{\mu \mu}^{qP}=\epsilon_{\tau \tau}^{qP}\sim 1$, $\epsilon_{ee}^{qP}=0$ and $\epsilon_{\alpha \beta}|_{\alpha \ne \beta}^{qP}=0$. For scattering experiments such as NuTeV with energy exchange of $q$ much higher than $m_X$, the amplitude of new effects will be suppressed by a factor of $(m_X/q)^2$  so the corresponding bounds can be easily satisfied.

The off-diagonal elements of the $\epsilon^{qP}$ matrix which violate lepton flavor can induce a significant effect on neutrino oscillation in matter. The DUNE and current NO$\nu$A experiments will be able to probe the values of these elements well below the present bound. In this paper, we explore possibilities of obtaining nonzero off-diagonal $\epsilon^{qP}$ elements
within $U(1)^\prime$ gauge models in which left-handed leptons sit in  a non-trivial two-component representation. Similarly to \cite{Me}, we will assume the new gauge boson is relatively light. As shown in \cite{Julian-Heeck}, the bound on $\tau \to Z^\prime \mu$ significantly constrains the new gauge coupling to leptons. To maintain sizeable non-standard effective couplings between leptons and quarks, we take $U(1)^\prime$ charges of the quarks to be much larger than those of leptons. This in turn implies that the contributions of quarks and leptons to $U(1)^\prime$ gauge anomalies should be canceled separately. The contribution to anomalies from leptons automatically cancel out. To cancel the anomaly from quarks, we introduce new generation(s) of leptons with appropriate
$U(1)^\prime$ charges. The lightest new lepton can play the role of the dark matter so we find a dark matter candidate as a bonus.

The paper is organized as follows. In section \ref{model}, we explore the possibilities to build a gauge interaction model with off-diagonal couplings to the leptons: $Z^\prime_\mu \bar{\nu}_\alpha \gamma^\mu \nu_\beta$ with $\alpha \ne \beta$. In section \ref{pheno}, we will specialize to the case of $\alpha=\mu$ and $\beta=\tau$ and discuss various observational bounds and constraints. In section \ref{con}, we summarize our conclusions.

\section{The Model\label{model}}
In this section we introduce a $U(1)^\prime$ gauge model that gives rise to an effective coupling of form  in Eq. (\ref{NSI}).  NSI involving the electron will affect electron-neutrino cross section in solar neutrino experiments such as Borexino and Super-Kamiokande. To avoid any deviation, following \cite{Maltoni} we set $f=q=\{ u,d \}$. To obtain an interaction between neutrinos and quarks, both leptons and quarks need to have nonzero $U(1)^\prime$ charges. If NSI with quarks is non-chiral ({\it i.e.,} if $\epsilon_{\alpha \beta}^{q L}\ne \epsilon_{\alpha \beta}^{q R}$), the rate of deuteron dissociation ($\nu+{\rm Deuteron} \to \nu+n+p$) used by the Sudbury Neutrino Observatory (SNO) to derive total solar neutrino flux will be affected. The agreement between the prediction of standard solar model and the total flux measured by SNO sets bounds on the $\epsilon_{\alpha \beta}^{q L}- \epsilon_{\alpha \beta}^{q R}$. To avoid such a constraint, we assume the coupling of $Z^\prime$ to quarks to be non-chiral $\epsilon_{\alpha \beta}^{q L}= \epsilon_{\alpha \beta}^{q R}$. In fact, the combination relevant for neutrino propagation in matter is the vectorial combination of $\epsilon_{\alpha \beta}^{q L}$ and $ \epsilon_{\alpha \beta}^{q R}$ $$\epsilon_{\alpha \beta}^q\equiv \epsilon_{\alpha \beta}^{q L}+ \epsilon_{\alpha \beta}^{q R}.$$
As emphasized in the introduction the aim of the present paper is to build a model giving rise to Lepton Flavor Violating (LFV) NSI of neutrinos with matter that can be probed at long baseline experiments. However, for simplicity we take the interaction to be flavor diagonal in the quark sector: That is we assume quarks are in the singlet representation of $U(1)^\prime$ and only obtain a phase under $U(1)^\prime$ transformation. Putting these together, we conclude that the $U(1)^\prime$ charges of quarks of each generation are the same. In other words, the $U(1)^\prime$ charges of quarks are given by \be \label{Bi} \eta_1 B_1+\eta_2 B_2+\eta_3 B_3\ee where $B_i$ is the Baryon number of the $i$ generation.
With this definition, the couplings of quark to $Z^\prime$ can be described as $\sum_{i=1}^3 \eta_i Z^\prime_\mu g^\prime (\bar{u}_i \gamma^\mu u_i +\bar{d}_i \gamma^\mu d_i)$ where $u_1=u$, $u_2=c$, $u_3=t$, $d_1=d$, $d_2=s$ and $d_3=b$.
If $\eta_i$ are not equal, in the quark mass basis, the $Z^\prime$ couplings can have nonzero off-diagonal elements. From the $D-\bar{D}$ mixing and Kaon physics, there are strong bounds on the 1-2 components of flavor changing neutral current. Similarly to \cite{Heeck} to avoid these bounds, we set $\eta_1=\eta_2$. {In case $\eta_3 \ne \eta_1=\eta_2$, in the mass basis, flavor changing couplings of form $ g^\prime(\eta_3-\eta_1) V_{tb}V_{ts}^* Z^\prime_\mu \bar{b}\gamma^\mu s$ and $ g^\prime(\eta_3-\eta_1) V_{tb}V_{td}^* Z^\prime_\mu \bar{b}\gamma^\mu d$ appear which can give rise to $b \to Z^\prime s$ and
$b \to Z^\prime d$. As is well-known because of the longitudinal components of $Z^\prime$, the rates of $b \to Z^\prime s$ and  $b \to Z^\prime d$  will be proportional to $m_b^3/m_{Z^\prime}^2$.  Since we want $m_{Z^\prime}\ll m_b$, this causes a huge enhancement. To avoid problems, we can set $\eta_1=\eta_2=\eta_3$; {\it i.e.,} we gauge Baryon number. }

Let us now discuss the transformation of leptons under $U(1)^\prime$ symmetry. To obtain off-diagonal components, we assume that two generations of left-handed doublets, $\tilde{L}_\alpha =(\tilde{\nu}_\alpha \ \tilde{l}^-_\alpha)$ and  $\tilde{L}_\beta =(\tilde{\nu}_\beta \ \tilde{l}^-_\beta)$, form a doublet of $U(1)^\prime$:
\be \tilde{L}\equiv\left( \begin{matrix} \tilde{L}_\alpha \cr \tilde{L}_\beta \end{matrix} \right) \stackrel{U(1)^\prime}{\longrightarrow} e^{i\zeta g^\prime\sigma_1 \alpha}\tilde{L} , \label{du}\ee
where $\sigma_1$ is the first Pauli matrix {and $\zeta/\eta_i $ gives the relative strength of coupling of leptons to that of quarks.}
 In this basis the coupling of the $Z^\prime$ boson to left-handed leptons will take the following form
$$ \zeta g^\prime Z^\prime_\mu \bar{\tilde{L}} \gamma^\mu \sigma_1\tilde{L}= \zeta g^\prime Z^\prime_\mu (\bar{\tilde{\nu}}_\alpha\gamma^\mu \nu_\beta+ \bar{\tilde{\nu}}_\beta\gamma^\mu \tilde{\nu}_\alpha+ \bar{\tilde{l}}_{L\alpha}\gamma^\mu \tilde{l}_{L\beta}+ \bar{\tilde{l}}_{L\beta}\gamma^\mu \tilde{l}_{L\alpha})\ .$$
Notice that the transformation as Eq. (\ref{du}) is equivalent to having two fields $(\tilde{L}_\alpha+\tilde{L}_\beta)/\sqrt{2}$ and  $(\tilde{L}_\alpha-\tilde{L}_\beta)/\sqrt{2}$ with opposite charges
\be \frac{\tilde{L}_\alpha+\tilde{L}_\beta}{\sqrt{2}}\to e^{i \alpha \zeta g^\prime} \frac{\tilde{L}_\alpha+\tilde{L}_\beta}{\sqrt{2}} \ \ {\rm and} \ \ \frac{\tilde{L}_\alpha-\tilde{L}_\beta}{\sqrt{2}}\to e^{-i \alpha \zeta g^\prime} \frac{\tilde{L}_\alpha-\tilde{L}_\beta}{\sqrt{2}}.\label{L-opp}\ee
We have used tilded symbols to emphasize that $|\tilde{l}_\alpha^-\rangle $ and $|\tilde{l}_\beta^-\rangle$ are not necessarily mass eigenstates. Denoting the charged leptons of definite mass  by ${l}_\alpha^-$ and ${l}_\beta^-$ and corresponding neutrinos with $\nu_\alpha$ and $\nu_\beta$, we can in general write:
\be L\equiv \left(\begin{matrix} L_\alpha \cr L_\beta  \end{matrix}\right)= \left(\begin{matrix} \cos \theta_L & -\sin \theta_L \cr \sin \theta_L & \cos \theta_L \end{matrix}\right)\tilde{L}\ .\label{tL}\ee
In the mass basis, the $U(1)^\prime$ gauge interaction will be of form
\be \zeta  g^\prime Z^\prime_\mu (\bar{L}_\alpha \ \bar{L}_\beta) \gamma^\mu  \left( \begin{matrix} -\sin 2\theta_L & \cos 2\theta_L \cr \cos 2\theta_L & \sin 2 \theta_L \end{matrix} \right) \left(\begin{matrix} L_\alpha \cr L_\beta\end{matrix}\right). \label{2tL}\ee
For $\theta_L \ne 0, \pi/2$, we shall have flavor conserving interactions, too.  There will be flavor violating gauge interactions for all values of $\theta_L$ within physical range except for $\theta_L =\pm \pi/4$.

Let us assume that the lepton generation denoted by $\gamma$ is singlet under $U(1)^\prime$. In the next section, we will identify $\gamma$ with first generation on which there are strong bounds. Notice that the contribution of $\tilde{L}$ (or equivalently $L$) to the $U(1)^\prime-SU(2)-SU(2)$ and  $U(1)^\prime-U(1)-U(1)$ anomalies automatically cancel out because $Tr(\sigma_1)=0$; however, there will be a contribution to the $U(1)-U(1)^\prime-U(1)^\prime$ anomaly from $\tilde{L}$. The contribution of quarks to the $U(1)^\prime-SU(3)-SU(3)$ anomaly is given by $+(-1/2)\times 2 \times Tr\{ \sigma_1\sigma_1 \}=-2$.

Notice that $\nu_R$ does not help to cancel the $U(1)-U(1)^\prime-U(1)^\prime$ anomaly since $\nu_R$ does not carry any hypercharge. To cancel the remaining
$U(1)-U(1)^\prime-U(1)^\prime$ anomaly from lepton side, we should assign an appropriate transformation to $l_{R \alpha}^-$ and $l_{R \beta}^-$. In the following, we suggest two solutions:
\begin{itemize}
\item  \textbf{  $l_{R \alpha}^-$ and $l_{R \beta}^-$ form a doublet of  $U(1)^\prime$:}
Let us define $R$ as a doublet under $U(1)^\prime$ which is formed from the right-handed charged leptons and transform under $U(1)^\prime$
as follows
\be \tilde{R}\equiv\left( \begin{matrix} \tilde{l}_{R\alpha}^- \cr \tilde{l}_{R\beta}^- \end{matrix} \right) \stackrel{U(1)^\prime}{\longrightarrow} e^{i\zeta g^\prime\sigma_1 \alpha}\tilde{R} .\label{R-doublet} \ee
{In other words, in the basis that the left-handed leptons have opposite $U(1)^\prime$ charges (see Eq. \ref{L-opp}), two right-handed leptons $(\tilde{l}_{R \alpha}^-+ \tilde{l}_{R \beta}^-)/\sqrt{2}$ and  $(\tilde{l}_{R \alpha}^-- \tilde{l}_{R \beta}^-)/\sqrt{2}$ have the ``same opposite" charges.}
It is straightforward to show that the contribution from $\tilde{R}$ to $U(1)-U(1)^\prime-U(1)^\prime$ cancels that from $\tilde{L}$. We can write the following Yukawa couplings invariant under electroweak as well as $U(1)^\prime$ symmetry
\be \label{bb}  b_0 \tilde{R}^\dagger H^\dagger \tilde{L}+   b_1 \tilde{R}^\dagger \sigma_1 H^\dagger \tilde{L},\ee
where $H$ is the SM Higgs and $b_0$ and $b_1$ are Yukawa couplings. After electroweak symmetry breaking these Yukawa interactions will induce the following mass matrix:
\be \tilde{R}^\dagger B \tilde{L}=( \tilde{l}_{R \alpha}^\dagger  \   \tilde{l}_{R \beta}^\dagger)\left( \begin{matrix} b_0 v & b_1 v \cr b_1 v & b_0 v \end{matrix} \right) \left( \begin{matrix}  \tilde{l}_{L \alpha}  \cr   \tilde{l}_{L \beta}\end{matrix} \right)\ee
Notice that the 11 and 22 components of the mass matrix $B$ are equal which implies the mixing angle  defined in Eq. (\ref{tL}) will be equal to $\pi/4$. This in turn implies that the coupling of $Z^\prime$ conserves flavor (see Eq. (\ref{2tL})). To solve this problem, we introduce a $2\times 2$  matrix $\Phi$ whose components are scalar fields, doublet under electroweak with the same hypercharge as that of the SM Higgs. Under $U(1)^\prime$, $\Phi$ transforms  as
 $$\Phi \to e^{i \zeta  g^\prime
\alpha \sigma_1} \Phi e^{-i \zeta  g^\prime \sigma_1\alpha}.$$
We can write a Yukawa coupling of form
$$ c_0 \tilde{R}^\dagger \Phi \tilde{L}.$$
The components of $\Phi$ can be heavy enough to avoid bounds but we can take $c_0\ll 1$ to obtain small contribution to the lepton masses. Taking $m_\beta^2 \gg m_\alpha^2$, we can write $$\cos 2\theta_L=\frac{2c_0v\left( b_0 ( \langle\Phi_{22}\rangle -\langle \Phi_{11}\rangle)+b_1 ( \langle\Phi_{12}\rangle -\langle \Phi_{21}\rangle)\right)}{m_\beta^2}$$
which can be in general nonzero, leading to flavor violating $Z^\prime$ coupling. Another issue is that the mass structure of charged leptons has to be hierarchical.
If we want the main contribution to heavier lepton to come from the vacuum expectation value  of the SM Higgs ({\it i.e.,} if $c_0 \langle \Phi \rangle \ll b_i v$),  $b_0$ and $b_1$ should be approximately equal to reconstruct the hierarchical mass pattern in the SM ({\it i.e.,} $m_\beta \gg m_\alpha$). Such an equality can be explained by discrete symmetry $\tilde{R}^\dagger \to \tilde{R}^\dagger\sigma_1$ and $\Phi \to \sigma_1 \Phi$. We can then write $b_0\simeq b_1 \simeq m_\beta /2v$. We can moreover write the couplings of  the Higgs to the charged leptons as
\be \frac{m_\beta}{2v} H (l^\dagger_{R\alpha}  \ l^\dagger_{R\beta}) \left( \begin{matrix} \cos \theta_R & -\sin \theta_R \cr \sin \theta_R & \cos\theta_R \end{matrix} \right) \left( \begin{matrix} 1 & 1 \cr 1 & 1 \end{matrix} \right)\left( \begin{matrix} \cos \theta_L & \sin \theta_L \cr -\sin \theta_L & \cos\theta_L \end{matrix} \right)  \left( \begin{matrix} l_{L\alpha}\cr l_{L\beta}\end{matrix} \right)\ee
where $\theta_R$ is the mixing angle relating $\tilde{R}$ to  the right-handed charged lepton mass eigenvectors. This can lead to LFV Higgs decay:
\be \frac{Br(H\to \bar{l}_{L\beta}  {l}_{R\alpha})}{Br(H\to \bar{l}_{L\beta}  {l}_{R\beta})}=\left( \frac{\sin \theta_R-\cos \theta_R} {\sin \theta_R+\cos \theta_R}\right)^2 \label{11}\ee
and
\be \frac{Br(H\to \bar{l}_{L\alpha}  {l}_{R\beta})}{Br(H\to \bar{l}_{L\beta}  {l}_{R\beta})}=\left( \frac{\sin \theta_L-\cos \theta_L} {\sin \theta_L+\cos \theta_L}\right)^2 .\label{12}\ee
Notice that for general complex $\Phi$ with Re($\langle\Phi_{21}\rangle) \ne$ Re$(\langle\Phi_{12}\rangle)$, the values of $\theta_R$ and $\theta_L$ are not the same.
\item {{
Right-handed charged leptons have opposite $U(1)^\prime$ charges.}} Let us now suppose:
 \be \label{R-singlet}\tilde{l}_{R\alpha} \stackrel{U(1)^\prime}{\longrightarrow} e^{i \zeta g^\prime q \alpha} \tilde{l}_{R\alpha}  \ \ \ {\rm and} \ \ \ \tilde{l}_{R\beta} \stackrel{U(1)^\prime}{\longrightarrow} e^{-i\zeta  g^\prime q \alpha} \tilde{l}_{R\beta}.\ee
With this charge assignment, the $U(1)^\prime-U(1)^\prime-U(1)^\prime$ anomaly automatically cancels out. The value of $q$ should be assigned such that the contribution from the right-handed leptons to the $U(1)-U(1)^\prime-U(1)^\prime$ anomaly
cancels that from the left-handed leptons.
That is $-2 q^2(-1)+2(-1/2) Tr(\sigma_1\sigma_1)=0$ which implies $q=1$.
To couple both $\tilde{l}_{R \alpha}$ and $\tilde{l}_{R\beta}$ to $\tilde{L}$, we need to introduce two scalar $U(1)^\prime$
doublets whose components are Higgs-like doublets under electroweak symmetry with hyper-charge equal to $+1/2$:
\be \Phi_1\equiv \left( \begin{matrix}\phi_{1\alpha} \cr \phi_{1\beta}\end{matrix} \right) \stackrel{U(1)^\prime}{\longrightarrow} e^{i \zeta g^\prime \alpha}e^{i \zeta g^\prime\sigma_1 \alpha} \Phi_1 \ \ \ {\rm and} \ \ \ \Phi_2\equiv \left( \begin{matrix} \phi_{2\alpha} \cr \phi_{2\beta}\end{matrix}  \right) \stackrel{U(1)^\prime}{\longrightarrow} e^{-i\zeta  g^\prime \alpha}e^{i \zeta g^\prime\sigma_1 \alpha} \Phi_2 \ee
With this field content, we can write Yukawa couplings of the following forms
\be \tilde{l}_{R\alpha}^\dagger \Phi_1^\dagger \tilde{L}, \ \tilde{l}_{R\alpha}^\dagger \Phi_1^\dagger \sigma_1 \tilde{L}, \ \tilde{l}_{R\beta}^\dagger \Phi_2^\dagger \tilde{L}, \ {\rm and} \ \tilde{l}_{R\beta}^\dagger \Phi_2^\dagger\sigma_1 \tilde{L}\ .\ee
After $\Phi_1$ and $\Phi_2$ develop vacuum expectation values, these terms give masses to charged leptons. In general,
$\theta_L\ne \pi/4$ so the  flavor violating $U(1)^\prime$ gauge couplings are obtained. The SM Higgs can mix with the neutral components of $\Phi_i$ so it can decay through this mixing to charged lepton pairs. However the rate of the Higgs decay into charged leptons will deviate from the SM prediction and will not be given by $(m_f/v)^2$. The decay rate of the SM Higgs into $\tau$ pair is now measured and found to be consistent with the SM prediction. This solution is not therefore suitable for the case that $\alpha$ or $\beta$ is identified with $\tau$.
\end{itemize}
In  both of two above cases, going to the lepton flavor basis arranged as $(\gamma \ , \alpha \ , \beta )$, the $\epsilon^{qP}$ matrix will be equal to
\be \label{fs}\epsilon^{u L} = \epsilon^{u R} = \epsilon^{d L} = \epsilon^{d R} =\frac{\zeta \eta_1 (g^\prime)^2} {m_{Z^\prime}^2}\frac{1}{2 \sqrt{2} G_F}\left( \begin{matrix} 0 & 0 & 0 \cr 0 & -\sin 2\theta_L & \cos 2 \theta_L \cr 0 & \cos 2\theta_L & \sin 2 \theta_L \end{matrix} \right) .\ee {Without loss of generality, we can set $\eta_1=1$.}

Notice that in neither of the above two solutions, right-handed neutrinos are required for anomaly cancelation. Although it is not the main subject of the present paper, let us provide an example to show how type I seesaw mechanism can be implemented within this model. Let us take the right-handed neutrinos singlet under $U(1)^\prime$, too.
Majorana mass matrix of right-handed neutrino as well as the Dirac mass term for $L_\gamma$ can be written as in the standard type I seesaw mechanism. To write the Dirac mass terms $\bar{\nu}_{R i} L_\alpha$ and $\bar{\nu}_{R i} L_\beta$, we however need a scalar doublet of $U(1)^\prime$ shown by $H_N$ whose components are doublets of electroweak and their hypercharge is equal to that of $L$.
We then obtain the desired Dirac mass terms via $\bar{\nu}_{R i} H_N^T c L$ and $\bar{\nu}_{R i} H_N^T c\sigma_1  L$. Vacuum expectation values of $H_N$ can be taken to be much smaller than that of the SM Higgs to avoid changing Yukawa coupling of $H$ to fermions $m_f /\langle H\rangle \simeq \sqrt{2} m_f/v$ where $v=246$ GeV). As shown
in \cite{Me}, components of $H_N$ can be made heavy despite small $\langle H_N^0\rangle$.

{As discussed before, the contribution of leptons to anomalies cancel. Taking $\eta_1=\eta_2=\eta_3$ in Eq. (\ref{Bi}), the quark sector  induces a contribution to the $U(1)^\prime-U(1)-U(1)$ and $U(1)^\prime-SU(2)-SU(2)$ anomalies but the $U(1)^\prime-SU(3)-SU(3)$ anomaly automatically cancels out. The anomaly can be canceled by adding new chiral degrees of freedom. On example is
 to introduce two generations of new leptons (with the same charges under SM gauge symmetry as those of leptons) with $U(1)^\prime$ charge equal to $-9/2 \eta_1$.\footnote{For cancelation of triangle anomalies, one generation of new leptons with a  charge of $-9 \eta_1$ will be enough. We suggest to add two generations of new leptons with two doublets to cancel the Witten anomaly, too.}
                 With such field content anomalies will be canceled. These chiral fermions, like quarks, can obtain mass by coupling to the SM Higgs. Perturbativity of their Yukawa coupling then implies an approximate upper bound of $O(600~{\rm GeV})$ on their mass.  These particles can be  produced at colliders via their electroweak interactions. The present lower bound on the mass of such  new charged charged leptons is about $O(100~{\rm GeV})$ \cite{pdg}.   The $U(1)^\prime$ symmetry prevent mixing between these new leptons and the SM leptons. Thus, the lightest new lepton, which can  correspond to the new neutrino, will be stable and can play the role of dark matter. Notice that this aspect of the scenario is only peripheral to the purpose of the present paper. We will not therefore elaborate on it further.}

{In the above discussion, we have introduced new scalars that have charges of $\zeta$ under $U(1)^\prime$ and transform as standard model Higgs under electroweak and develop Vacuum Expectation Value (VEV). Their VEV will both induce a mass for $Z^\prime$ and  mixings between $Z^\prime$ with $Z$ (but not with $\gamma$). These mass parameters are of order of $g^\prime \zeta$ times the VEV of these new scalars. Since the VEV of these new scalars are taken to be smaller than $\langle H\rangle$, the mass terms created by their VEV will be smaller than $g^\prime \zeta \langle H\rangle$. As we shall see in the next section, $g^\prime \zeta$ is constrained to be smaller than $3\times 10^{-9} (m_{Z^\prime}/ 10~{\rm MeV})$. These mass terms are therefore much smaller than $m_{Z^\prime}$ and can be safely neglected.  To explain the mass of $Z^\prime$, we can either invoke the St\" uckelberg mechanism or introduce a scalar $(S)$ singlet under SM gauge symmetry and a $U(1)^\prime$ charge of $\zeta^\prime$. The mass of  $Z^\prime$ will be given by $g^\prime \zeta^\prime \langle S\rangle \sim {\rm few} \times 10 ~{\rm MeV}$. Notice that since $g^\prime \zeta^\prime$ can be arbitrarily small, $\langle S\rangle$ can be made large. The new scalar can be much heavier than $Z^\prime$. }
\section{Phenomenological Implications\label{pheno}}
{
In this section, we discuss the observational effects of the model presented in the previous section and discuss the bounds on its parameters from various observations and experiments. The observational imprint of the model depends on the decay modes of $Z^\prime$. Since we take $m_{Z^\prime}<2 m_\mu$, it cannot decay into muon and tau lepton pairs but $Z^\prime$ can decay into $\nu_\alpha \bar{\nu}_\alpha$, $\nu_\beta \bar{\nu}_\beta$ as well as $\nu_\alpha \bar{\nu}_\beta$ and $\nu_\beta \bar{\nu}_\alpha$. If $\alpha$ is identified with $e$, $Z^\prime$ can decay into $e^-e^+$, too. Moreover if $\alpha =e$, $\beta=\mu$ and $m_{Z^\prime}>m_\mu$, we can have $Z^\prime \to \mu^-e^+$ and $Z^\prime \to e^-\mu^+$. Although in our model quarks also couple to $Z^\prime$ (and as we shall see, with a coupling much larger  than those of leptons), as far as $m_{Z^\prime}<m_\pi$, $Z^\prime$ will not have hadronic decay modes because the lightest hadrons ({\it i.e.,} pions) are heavier than $Z^\prime$. This is the famous mass gap problem which appears in the confined regime of strongly interacting theories. At first sight, it seems that via quark loops $Z^\prime$ and photon mix which can give rise to $Z^\prime \to e^- e^+$ even for the $\alpha=\mu$ and $\beta =\tau$ case with no tree level coupling between the electron and $Z^\prime$. However for energy scale below the QCD scale ($\sim {\rm few} \times 100$~MeV) instead of quarks, hadrons should propagate in the loops. This is well-known in the calculation of hadronic loop for vacuum polarization of the photon which is needed
for precise calculation of $(g-2)_\mu$ (see, for example, \cite{g-2}.) On the other hand, mesons (having zero  baryon number) are neutral under $U(1)^\prime$ so they cannot mix $Z^\prime$ and the photon. The lightest hadron charged under both $U(1)_{em}$ and $U(1)^\prime$ is the proton which is much heavier than the scale that we are interested in and therefore is decoupled from low energy physics. As a result, for $\alpha =\mu$ and $\beta=\tau$, the only available $Z^\prime$ decay mode is into neutrino pairs $\nu_\mu \bar{\nu}_\mu$, $\nu_\tau \bar{\nu}_\tau$, $\nu_\mu \bar{\nu}_\tau$ and $\nu_\tau \bar{\nu}_\mu$.}

{If $Z^\prime$ decays into $ e^-e^+$, it can be traced in the beam dump fixed target experiments \cite{E137,E141,E774}, for a given $m_{Z^\prime}$, these experiments rule out values of coupling between an upper bound and a lower bound. The $Z^\prime$ production in these experiments is through their couplings to the quarks. If the coupling to $e^-e^+$ is higher than the upper bound, the $Z^\prime$ decay takes place in the dump and the produced $e^-e^+$ will not be registered. On the other hand, if the couplings to quarks and leptons are too small, the rates of the $Z^\prime$ production and of the subsequent decay into $e^-e^+$ will be too low to have an observable effect. The beam dump bounds are derived and shown in \cite{E137,E141,E774} for the models that the $U(1)^\prime$ charges of the electron and the quarks are equal.  The upper limit of the excluded region, determined by  $\Gamma (Z^\prime \to e^- e^+)$, can be readily interpreted as the upper limit of the excluded region of $\zeta g^\prime \sin 2 \theta_L$ in our model in case that $\alpha$ is identified with the electron. This bound can be further improved by the SHiP experiment \cite{Ship}. The lower limit of the excluded region in our case should be however stronger than  \cite{E137,E141,E774} because the $Z^\prime$ production is determined by quark charge which is fixed to $+1$ ({\it cf.} lepton charges are suppressed by $\zeta$). If $\alpha, \beta \ne e$, $Z^\prime$ does not decay to $e^- e^+$ so the beam dump bounds do not apply.
}

The bounds on the couplings of a new $U(1)^\prime$ gauge boson to quarks  are mostly derived by looking for the $e^- e^+$ pair from the decay of $Z^\prime$ produced in various density frontier experiments. Most importantly Ref \cite{WASA} sets a bound $g^\prime <10^{-3}$ for $m_{Z^\prime}\sim 20$~MeV from $\pi^0\to \gamma Z^\prime$ and subsequently $Z^\prime \to e^-e^+$. As we show below, we can find a comparable bound for the case that $Z^\prime$ decays into $\nu \bar{\nu}$ from Br($\pi^0 \to \nu \bar{\nu} \gamma) <6 \times 10^{-4}$ \cite{pdg}.  We expect
\be \label{2/3}{\rm Br}(\pi^0\to \gamma Z^\prime)=2 \left(\frac{g^\prime (2/3-(-1/3))}{e [(2/3)^2-(-1/3)^2]}\right)^2\left(1-\frac{m_{Z^\prime}^2}{m_{\pi^0}^2}\right)^3<6\times 10^{-4}.\ee
A similar formula can be found in \cite{Gninenko} for the case of a $U(1)^\prime$ gauge boson mixed with the photon. The factor of 2 reflects the fact that unlike the case of $\pi^0 \to \gamma \gamma$,    the final particles in  $\pi^0 \to \gamma Z^\prime$ are  distinguishable. Taking ${\rm Br}(\pi^0\to \gamma Z^\prime)\leq Br(\pi^0\to \gamma \nu \bar{\nu})< 6\times 10^{-6}$, we find \be {\rm for}~~ m_{Z^\prime}<m_{\pi^0}=135~{\rm MeV}~~~~~~~~~  g^\prime <2\times 10^{-3} \label{uppp}.\ee  For $m_{Z^\prime}>135$ MeV, $\pi^0$ cannot decay into $Z^\prime \gamma$ so no bound can be set from pion decay on $g^\prime$. To our best knowledge, all the bounds set on $g^\prime$ for $135~{\rm MeV}<m_{Z^\prime}<200~{\rm MeV}$ are based on searching for leptons from the $Z^\prime$ decay \cite{HADES,A1,APEX,NA48/2} which do not apply to our case. For $m_{Z^\prime}>200$ MeV when the $Z^\prime$ decay into $\mu^-\mu^+$ becomes possible, there are stronger bounds from BaBar \cite{babar} and KLOE-2 \cite{KLOE2}. Notice that for $m_\pi <m_{Z^\prime}<200$~MeV, the range of the $U(1)^\prime$ interaction between nucleons is comparable to that of strong interactions so as long as $(g^\prime)^2/4\pi <\alpha$, its effects will be too small to be discerned in the presence of strong interactions. Throughout this section, we assume that $g^\prime$ saturates this bound so that we can obtain a sizeable $\epsilon_{\alpha \beta}^{qP}$.
In this mass range the $Z^\prime$ can decay into $\pi^0 \gamma$ but since decay takes place inside the dump, the beam dump experiments cannot identify it.

With such large $g^\prime$, the $Z^\prime$ particles can be produced at supernova core via $N+N \to N+N+Z^\prime$. The $Z^\prime$ particles will thermalize in the supernova core via interactions with nucleons with a mean free path smaller than 1~cm in the supernova core and eventually decay into standard model particles. The $Z^\prime$ production and decay can take place  outside neutrinosphere, too. It will be interesting to study its possible effects on supernova evolution and shock revival but such an analysis is beyond the scope of the present paper.

{In the present model, we have a LFV coupling of form $$g^\prime \zeta Z_\mu^\prime\left(\cos 2  \theta_L (\bar{l}_{L \alpha}\gamma^\mu l_{L\beta} +\bar{l}_{L\beta}\gamma^\mu l_{L\alpha})+\cos 2  \theta_R (\bar{l}_{R \alpha}\gamma^\mu l_{R\beta} +\bar{l}_{R\beta}\gamma^\mu l_{R\alpha})\right)$$ (see Eq. (\ref{2tL})) which opens a new decay mode $l_\beta \to Z^\prime l_\alpha$ \cite{Julian-Heeck}
with \be \label{LFVtau} \Gamma (l_\beta^- \to Z^\prime l_\alpha^-)=\frac{(g^{\prime})^2\zeta^2 }{32 \pi}\frac{m_{l_\beta}^3}{m_{Z^\prime}^2} (\cos^2 2\theta_L+\cos^2 2\theta_R) \ee
where we have neglected the terms suppressed by $(m_{Z^\prime}/m_{l_\beta})^2 , (m_{l_\alpha}/m_{l_\beta})^2\ll 1$. Let us discuss the implications of the bounds on $\Gamma (l_\beta^- \to Z^\prime l_\alpha^-)$ for different flavor compositions $\alpha$ and $\beta$ one by one. }

 {\it{Case I, $\beta=\mu$  and $\alpha=e$:}}
 If we identify $\beta=\mu$ and $\alpha=e$ (and therefore $\gamma=\tau$) we will have $\mu \to e Z^\prime$ and subsequent decay of $Z^\prime \to e \bar{e}$ will produce a signal of $\mu \to eee$ on which there is a strong bound ${\rm Br}(\mu \to eee)<10^{-12}$ \cite{pdg}. For $m_{Z^\prime}\sim 10$~MeV, this bound translates into $g^\prime \zeta ((\cos^2 2\theta_L+\cos^2 2\theta_R)/2)^{1/2}<10^{-13}$ which is so strong that kills any hope for sizeable $\epsilon_{\mu e}$. However for $m_{Z^\prime}>m_\mu$, $\mu \to Z^\prime e$ is not possible but $\mu \to eee$ can take place via a tree level diagram in which virtual $Z^\prime$ is exchanged with $\Gamma(\mu \to eee)\sim (\zeta g^\prime)^4 \cos^2 2\theta_L\sin^2 2\theta_L m_\mu^5/(100\pi^3m_{Z^\prime}^4)$.  From the upper bound on ${\rm Br}(\mu \to eee)$, we find $g^\prime \zeta \sqrt{\sin 2\theta_L\cos 2\theta_L}<6 \times 10^{-7} m_{Z^\prime}/(150~{\rm MeV}).$ For $m_{Z^\prime}\sim 150$~MeV and $Z^\prime \to e^+e^-$, the KLOE experiment finds $g^\prime<8 \times 10^{-4}$ by studying  $\phi \to Z^\prime \eta$ and subsequently $Z^\prime \to e^-e^+$. Putting these two bounds
 together we find $\epsilon_{e\mu}^{qP}<5 \times 10^{-4}$.  We will not investigate this case further.

   {\it{Case II, $\beta=\tau$  and $\alpha=e$:}} Let us now consider the case of $\alpha =e$ and $\beta=\tau$. The bound ${\rm Br}(\tau \to e Z^\prime)<2.7\times 10^{-3}$ \cite{pdg} sets the bound $$g^\prime \zeta (\cos^2 2\theta_L +\cos^2 2\theta_R)^{1/2} (\frac{m_{Z^\prime}}{10~{\rm MeV}})< 2\times 10^{-9}$$ which along with $g^\prime <2\times 10^{-3}$ gives $\epsilon_{\tau e}^{qP}<1.5 \times 10^{-3}$. The beam dump experiment E137 \cite{E137} rules out $Z^\prime$ with
 $m_{Z^\prime}\sim 10$~MeV with a coupling larger than $3 \times 10^{-8}$ to quarks and the electron. In our case, since the coupling to quarks (determining the production of $Z^\prime$ in this experiment) is much larger than the coupling to the leptons, the bound from E137 should be reconsidered. Performing this analysis is beyond the scope of the present paper.

   {\it{Case III, $\beta=\tau$  and $\alpha=\mu$}}
 As shown in \cite{Julian-Heeck}, the present bound on the branching ratio of this mode, ${\rm Br}(\tau \to Z^\prime \mu)<5 \times 10^{-3} $ \cite{pdg}, sets a severe bound on \be \label{tauz}\zeta  (\cos^ 2 2\theta_L+\cos^2 2 \theta_R)^{1/2}<3\times 10^{-9}(\frac{1}{g^\prime})(\frac{m_{Z^\prime}}{10~{\rm MeV}}).\ee The reason why the bound is so strong is the fact that $\Gamma (\tau \to Z^\prime \mu)$ is enhanced by
$(m_\tau/m_{Z^\prime})^2$ which is the famous factor due to the production of longitudinally polarized vector boson. At $m_{Z^\prime}<m_\pi $ , the bound on the coupling of quarks to $Z^\prime$ is also rather strong (see Eq. (\ref{uppp})). Taking $g^\prime \sim 10^{-3}$, $\zeta \sim 3 \times 10^{-5}$, $\cos 2 \theta_L\sim 0.1$ and
for  $m_{Z^\prime} \sim 10$~MeV, from Eq. (\ref{fs}) we find $$\epsilon_{\mu \tau}^{qP}\sim 10^{-3}~~~~~{\rm and} ~~~~~
|\epsilon_{\mu \mu}^{qP} -\epsilon_{\tau \tau}^{qP}|\sim 10^{-3}/\cos 2\theta_L\sim 0.01.$$ In this corer of parameter range, the model resembles millicharged models in which the $U(1)^\prime$ charge of one sector (in this case leptons) is several orders of magnitude smaller than the $U(1)^\prime$ charge of the other sector ({\it i.e.,} quarks).
As mentioned before, if $m_{Z^\prime}>m_\pi$, the bound on $g^\prime$ for this case is considerably relaxed. Taking $m_{Z^\prime}=140$ MeV, $g^\prime \sim 0.1$, $\zeta\sim 3\times 10^{-6}$ (and therefore still satisfying the bound from $\tau \to Z^\prime \mu$), we find
\be \label{WOW} \epsilon_{\tau \mu}^{qP} \sim 5 \times 10^{-3} \ \ \ {\rm and} \ \ \ \epsilon_{\mu \mu}^{qP}-\epsilon_{\tau \tau}^{qP} \sim 5 \times 10^{-3}/\cos 2\theta_L\sim 0.05. \ee These values of $\epsilon^{qP}_{\mu \tau}$ and $\epsilon_{\mu \mu}^{qP}-\epsilon_{\tau \tau}^{qP}$ are within the reach of the planned long baseline neutrino experiments \cite{deGouvea:2015ndi,Coloma:2015kiu}.

As discussed above, we can obtain sizeable $\epsilon_{\alpha \beta}^{qP}$  only for the case of $\epsilon_{\mu \tau}^{qP}$. We therefore focus in this case. Using
the terminology of the last section, we take $\gamma=e$, $\alpha=\mu$ and $\beta=\tau$.
We studied observational effects and various bounds from Borexino on neutrino-electron {interactions}~\cite{Bellini:2011rx}, NuTeV neutrino-nucleus scattering~\cite{Davidson:2003ha}, cosmic neutrino absorption at IceCube \cite{Aartsen:2015yva,Ioka:2014kca,Ng:2014pca,Ibe:2014pja,Blum:2014ewa,Cherry:2014xra,mm,DiFranzo:2015qea,Araki:2015mya} and neutrino trident production from CCFR \cite{Mishra:1991bv,Altmannshofer:2014pba}. We have found that lepton couplings suppressed by $\zeta\sim 10^{-5}$, the effects on these observations and experiments will be negligible.
The relevant {bounds} (also displayed in Fig 1) are the following:
  \begin{enumerate}

  \item {Relativistic Degrees of Freedom}: New light degrees of freedom can impact cosmology by changing the relativistic energy density and thus the expansion rate.
    The cosmological impact of light degrees of freedom is parameterized by the $N_{{\rm eff}}$ parameter, defined as $\rho_{{\rm rad}} = \rho_{\gamma} \left[1+ \frac{7}{8}\left(\frac{4}{11}\right)^{4/3}N_{{\rm eff}}\right]$, where $\rho_{{\rm rad}}$ is the total radiation energy density and $\rho_{\gamma}$ is the photon energy density.  Minimally we will need $m_{Z'} > 0.1$ MeV in order not to contribute all of its entropy directly to $N_{{\rm eff}}$ (which would result in $\Delta N_{{\rm eff}} \simeq 1.71$). This would considerably exceed the BBN constraint, of $\Delta N_{{\rm eff}} = 1.13$ (68$\%$ C.L.)~\cite{Steigman:2012ve}.  However, even if the $Z'$ is not directly contributing to $N_{{\rm eff}}$ as radiation itself, the entropy it transfers to $\nu_{\mu}$ and $\nu_{\tau}$ can raise their temperatures compared to the standard model case. We use the conservation of entropy to compute the temperatures of the $\mu$ and $\tau$ type neutrinos after the $\nu_{e}$ neutrinos undergo electroweak decoupling at $T_{{\rm SM},\nu} \simeq 1~{\rm MeV}$. We find that this rules out $m_{Z'} < 5~{\rm MeV}$ in good agreement with the results of~\cite{mm}. {Notice that for $m_{Z^\prime}>5$~MeV and $g^\prime \zeta>4 \times 10^{-11}$, the lifetime of $Z^\prime$ will be shorter than 1 sec.}

         \item {Neutrino oscillation data: current limits and prospect of DUNE and NO$\nu$A}: The long baseline of the DUNE and No$\nu$A experiments make them a natural setup to look for NSI~
\cite{Friedland:2012tq,deGouvea:2015ndi,Coloma:2015kiu,Masud:2015xva}. The combinations that are relevant for neutrino oscillation in matter are $\epsilon_{\alpha \beta} =\sum_{f\in \{ e,u,d \}} (n_f/n_e) \epsilon_{\alpha \beta}^f$
where $n_{f}$ is the number density of fermion $f$. In the crust of the Earth, $n_u/n_e=n_d/n_e=3$ so within our model, we can write $\epsilon_{\alpha \beta}\simeq 6 \epsilon_{\alpha \beta}^u= 6\epsilon_{\alpha \beta}^d$. Moreover by $\epsilon_{\alpha \alpha}\to \epsilon_{\alpha \alpha}-\epsilon_{\tau \tau}I_{3\times 3}$, the neutrino oscillation pattern will remain unaltered.  Following
\cite{Coloma:2015kiu}, let us
 redefine the diagonal components as $\tilde{\epsilon}_{ee} \equiv \epsilon_{ee} - \epsilon_{\tau \tau}$ and $\tilde{\epsilon}_{\mu \mu} \equiv \epsilon_{\mu \mu} - \epsilon_{\tau \tau}$. In our model these parameters are correlated and can be written as $\tilde{\epsilon}_{\mu \mu} = 2A \sin (2\theta_{L})$ and $\epsilon_{\mu \tau} = A \cos(2 \theta_{L})$ where
\be
A = \frac{6 \zeta g'^{2}}{2\sqrt{2} m_{Z'}^{2} G_{F}}.
\ee
         {Bounds on $\epsilon_{\mu \tau}$ and $\tilde{\epsilon}_{\mu \mu}$ can be translated into a bound on $A = \sqrt{\epsilon^{2}_{\mu \tau} + \tilde{\epsilon}^{2}_{\mu \mu} /4}$. DUNE (NO$\nu$A) sensitivity has been estimated to be $\epsilon_{\mu \tau} \sim 0.021~(0.031)$ and $\tilde{\epsilon}_{\mu \mu} \sim 0.1~(0.15)$~\cite{Coloma:2015kiu} which is equivalent to sensitivity to $A$ down to 0.054 (0.081). This can be translated into a bound on $\sqrt{\zeta} g^\prime/m_{Z^\prime}$. Notice that the bound on $\sqrt{\zeta}g^\prime/m_{Z^\prime}$ is given by the  root of $A$. As a result, the projected sensitivities of NO$\nu$A and DUNE to $ \sqrt{\zeta}g^\prime/m_{Z^\prime}$ are very close to each other. That is why we are showing them collectively in Fig 1 with a single line. To draw this line we have set $\zeta =4 \times 10^{-5}$.
         Long baseline experiments such as NO$\nu$A or DUNE can in principle disentangle $\epsilon_{\mu \tau}$ and $\tilde{\epsilon}_{\mu \mu}$. That means they can determine not only the value of  $A$ but also that of $\theta_L$. Fig \ref{fig1} also shows the constraint from the current oscillation bounds summarized in Eq. (\ref{current-bounds}) (which is equivalent to $A<0.108$) taking  $\zeta=3\times 10^{-6}$. This bound can be written as \be g^\prime<  \frac{7.7 \times 10^{-6}}{\sqrt{\zeta}} \frac{m_{Z^\prime}}{10~{\rm MeV}}\left( \frac{A}{0.0108} \right)^{1/2}.\label{g-prime-bound}\ee
         As we showed in Eq.~(\ref{tauz}), for a given $g^\prime/m_{Z^\prime}$, the bound on ${\rm Br}(\tau \to Z^\prime \mu)$ can be interpreted as an upper bound on $\zeta \cos 2 \theta_L$. Setting $\zeta \cos 2\theta_L$ equal to this bound  and asking $\epsilon_{\mu \tau}=6 \epsilon_{\mu \tau}^u$ to be larger than certain value we find the following lower bound on $g^\prime$:
         \be \label{lowerB}g^\prime >3.5 \times 10^{-4} \left( \frac{m_{Z^\prime}}{{\rm MeV}} \right)\frac{\epsilon_{\mu \tau}}{0.03}\ee
         where $\epsilon_{\mu \tau}=6 \epsilon_{\mu \tau}^u$. In Fig. 1, we show this lower bound for $\epsilon_{\mu \tau}=6 \epsilon_{\mu \tau}^u=0.06$ (present bound) and
         $\epsilon_{\mu \tau}=6 \epsilon_{\mu \tau}^u=0.021$ (DUNE reach). As seen from the figure for $5~{\rm MeV}<m_{Z^\prime}<9~{\rm MeV}$ and for $m_{Z^\prime}>130$ MeV, we can obtain values of $\epsilon_{\mu \tau}$ observable at DUNE. Moreover, for $5~{\rm MeV}<m_{Z^\prime}<20~{\rm MeV}$ and $m_{Z^\prime}>110$ MeV, we can have $\epsilon_{\mu \mu}-\epsilon_{\tau \tau}$ large enough to be discerned at DUNE.}


\end{enumerate}
\begin{figure}[t]
\begin{center}
 \includegraphics[width=.5\textwidth]{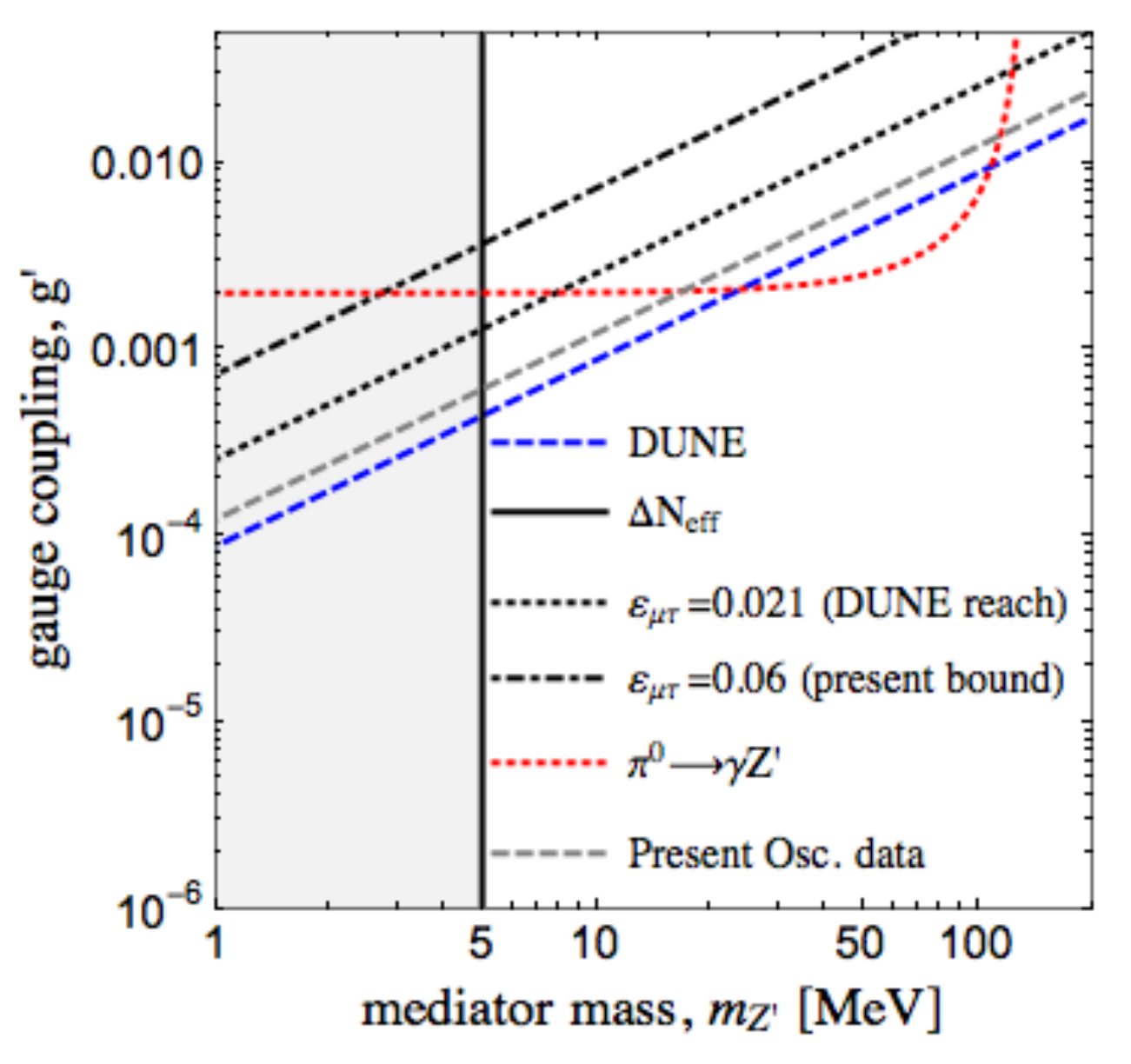}
\caption{{Here we summarize the constraints on the model. 
The dashed  blue and gray lines show the upper bounds from respectively present oscillation data and DUNE experiment incorporating information both on $\epsilon_{\mu \mu}-\epsilon_{\tau \tau}$ and $\epsilon_{\mu \tau}$ (see Eq. \ref{g-prime-bound}).  The {{dotted}} and dot-dashed black lines show the bound from only $\epsilon_{\mu \tau}$ for $\epsilon_{\mu\tau}=6 \epsilon_{\mu \tau}^u=0.021$ and
$\epsilon_{\mu\tau}=6 \epsilon_{\mu \tau}^u=0.06$, respectively (see Eq. \ref{lowerB}). The red curves shows the upper bound from $\pi^0 \to \gamma Z^\prime$ (see Eq. \ref{2/3}).  
Modifications to the effective number of relativistic degrees of freedom constrain the mass of the mediator to be $\gtrsim 5$ MeV~\cite{mm}. For additional details see the main body of the text.}}
\label{fig1}
\end{center}
\end{figure}

{For $200~{\rm MeV}\lesssim m_{Z^\prime}\lesssim 10$~GeV, B-factories impose  relatively strong bounds on $g^\prime$ particularly from $\Upsilon \to \gamma Z^\prime$ \cite{Essig:2009nc} and from $\Upsilon \to {\rm invisible}$ \cite{Graesser:2011vj}. In this mass range,  the $Z^\prime$ can decay into  $\mu \bar{\mu}$, hadrons, and $\nu \bar{\nu}$ pairs. For $m_{Z^\prime}>500$~MeV, this bound starts to become serious as it pushes $\epsilon$ to values lower than the present bounds in Eq. (\ref{current-bounds}). Throughout our discussion, we have assumed a  gauge boson mass, $m_{Z^\prime}<200$~MeV. In this range,  $\Upsilon \to \gamma Z^\prime$  and  $\Upsilon \to  Z^\prime Z^\prime$  are negligible. This can be understood as a consequence of the fact that in the limit of $m_{Z^\prime}\to 0$, according to the Landau-Yang theorem, the spin one $\Upsilon$ particle cannot decay into two massless or very light vector bosons.}

{In the following, we will discuss consequences of having  lepton flavor violating gauge couplings within the present model.
Most of these bounds come from processes of type $\tau \to \mu +f_1+\bar{f}_2$ where $f_1$ and $\bar{f}_2$ are final fermions and can be any of the pairs $u \bar{u}$, $d \bar{d}$, $\nu_\mu \bar{\nu}_\mu$,  $\nu_\tau \bar{\nu}_\tau$, $\nu_\tau \bar{\nu}_\mu$ and $\nu_\mu \bar{\nu}_\tau$.
The process takes place via the exchange of a virtual $Z^\prime$ and the amplitude can be written as
\be    \frac{\zeta g^\prime}{2} (\bar{\mu}[(\cos 2\theta_L+\cos 2 \theta_R)+ (-\cos 2\theta_L+\cos 2 \theta_R)\gamma^5]\gamma^\mu \tau)\frac{\eta_{\mu \nu}-q_\mu q_\nu/m_{Z^\prime}^2}{q^2-m_{Z^\prime}^2}( \bar{f}_1 (a_V+a_A\gamma^5)\gamma^\nu f_2)\ee
where,  for $f_1=f_2=u,d$, $a_V=g^\prime$ and $a_A=0$, for $f_1=f_2=\mu$,  $a_V=\zeta g^\prime (-\sin 2\theta_L-\sin 2\theta_R)/2$ and $a_A=\zeta g^\prime (\sin 2\theta_L-\sin 2\theta_R)/2$, for $f_1=\nu_\mu$ and $f_2=\nu_\tau$ as well as for $f_1=\nu_\tau$ and $f_2=\nu_\mu$, $a_V=-a_A=g^\prime \zeta \cos 2 \theta_L$, for $f_1=f_2=\nu_\mu$, $a_V=-a_A=-g^\prime \zeta \sin 2 \theta_L$ and finally for  $f_1=f_2=\nu_\tau$, $a_V=-a_A=g^\prime \zeta \sin 2\theta_L$. Using Dirac equation for $f_1$ and $f_2$ we can write $q_\nu \bar{f}_1 \gamma^\nu f_2=m_{f_1}-m_{f_2}=0$ and $q_\nu \bar{f}_1 \gamma^\nu\gamma^5 f_2=(m_{f_1}+m_{f_2})\ll m_\tau$. As a result, when $f_1$ and $f_2$ correspond to quarks (for which $a_A=0$) or to neutrinos (for which $m_{f_{1,2}}\to 0$), we can drop the terms proportional to $1/m_{Z^\prime}^2$ in the propagator of $Z^\prime$ which comes from longitudinal degrees of freedom. The amplitude for $\tau \to \mu q \bar{q}$ will be given by $\zeta (g^\prime m_\tau)^2/(q^2-m_{Z^\prime}^2)$ multiplied by  a function of sines and cosines of $\theta_L$ and $\theta_R$. Similarly, for neutrinos it will be given by $(g^\prime \zeta m_\tau)^2/(q^2-m_{Z^\prime})^2$ again  multiplied by  a function of sines and cosines of $\theta_L$ and $\theta_R$. For $\tau \to \mu \mu \mu$ longitudinal components of the propagator gives rise to an amplitude given by $(g^\prime \zeta )^2(\sin 2 \theta_R-\sin 2 \theta_L)[m_\tau^2/(q^2-m_{Z^\prime}^2)](m_\mu m_\tau/m_{Z^\prime}^2)$ again multiplied by a function of sines and cosines of $\theta_L$ and $\theta_R$. When $4 m_\mu^2 \sim q^2 \ll m_\tau^2$, the amplitude will be enhanced. As a result, the integration of $|M|^2$ over the phase space of the three final particles will yield a $\log (m_\tau^2/4m_\mu^2) \sim 4$ enhancement which does not change the order of magnitude of the decay rate.  }
\begin{itemize}
\item \textbf{Contribution to $\Gamma \left( \tau^- \to \mu^- {\nu}_\tau \bar{\nu}_\mu \right):$} The amplitude of the contribution from $Z^\prime$ to this process has to be added to the SM prediction for the same process. The relative size of the new amplitude is given by $(\zeta^2 g^{\prime 2}\cos^2 2 \theta_L/m_\tau^2)G_F^{-1}\sim 3\times 10^{-11} (\zeta g^\prime \cos 2\theta_L /3\times 10^{-8})^2 $. This is far below the precision of the measurement of {Br$(\tau \to \mu\bar{\nu}_\mu \nu_\tau) = \left(17.41 \pm 0.04\right) \%$~\cite{pdg}}. The contributions to $\tau \to \mu \bar{\nu}_\mu \nu_\mu$ or
    $\tau \to \mu \bar{\nu}_\tau \nu_\tau$ have no counterpart from SM. The relative branching ratio $[\Gamma (\tau \to \mu +\bar{\nu}_\mu \nu_\mu)+
    \Gamma (\tau \to \mu +\bar{\nu}_\tau \nu_\tau)]/ \Gamma (\tau \to \mu +\bar{\nu}_\mu \nu_\tau)$ will be therefore given by $(\zeta^4 g^{\prime 4} \cos^2 2\theta_L/m_\tau^4)G_F^{-2} \sim 8\times 10^{-20}(\cos^2 2\theta_L/0.01) (\zeta g^\prime /3\times 10^{-7})^4$ which is completely negligible.
    \item  \textbf{ LFV rare decay $\tau \to \mu \mu \mu$:}
    We can write
    \begin{equation}\hspace*{-4cm}\frac{\Gamma (\tau \to \mu \mu \mu)}{\Gamma (\tau \to \mu \nu_\tau \bar{\nu}_\mu)}\sim \frac{\zeta^4 g^{\prime 4} (\cos^2 2\theta_L+\cos^2 2\theta_R)
    (\sin 2\theta_L-\sin 2\theta_R)^2 }{m_\tau^4 G_F^2}\frac{m_\tau^2 m_\mu^2}{m_{Z^\prime}^4}\ee $$\hspace*{1.5 cm}\sim 2 \times 10^{-20} \frac{\cos^2 2\theta_L+\cos^2 2\theta_R}{0.01}  (\sin 2\theta_L-\sin 2\theta_R)^2 \left(\frac{\zeta g^\prime}{3\times 10^{-7}}\right)^4\left( \frac{150~{\rm MeV}}{m_{Z^\prime}}\right)^4$$
which is well below the present bound from ${\rm BR}(\tau \rightarrow \mu \mu \mu) < 2.1 \times 10^{-8}$~\cite{pdg,Hayasaka:2010np}.
            \item \textbf{ LFV rare decay  $\tau \to \mu q \bar{q}$:}
            Similarly to the previous case we can write
            \be \frac{\Gamma (\tau \to \mu +{\rm hadrons})} {\Gamma (\tau \to \nu_\tau +{\rm hadrons})}\sim
            \frac{ 2g^{\prime 4} (\cos^2 2\theta_L +\cos^2 2\theta_R)\zeta^2}{m_\tau^4 G_F^2}\sim (\epsilon_{\mu\tau}^{uL})^2 \frac{m_{Z^\prime}^4}{m_\tau^4}(1+\frac{\cos^2 2\theta_R}{\cos^2 2\theta_L})\ee $$\hspace*{-2cm} \sim 10^{-9} (\epsilon_{\mu \tau}^{u L})^2 (1+\frac{\cos^2 2\theta_R}{\cos^2 2\theta_L})$$
            which for $\epsilon_{\mu \tau}^{u L}\sim 0.01$ is well below the present bound from {${\rm BR}(\tau \rightarrow \mu \pi^{0}) < 1.1 \times 10^{-7}$~\cite{pdg}.}
        \item \textbf{One-loop correction to $\tau \to \mu \gamma$}: For $m_\tau \gg m_{Z^\prime}$, one loop-level contribution to $\tau \to \mu \gamma$ can be estimated as
        $$\Gamma (\tau \to \mu \gamma)\sim \frac{e^2g^{\prime 4} \zeta^4 }{8^3 \pi^5} m_\tau\cos^2 2\theta_L\sin^2 2\theta_L$$ which can be translated into ${\rm BR}(\tau \to \mu \gamma) \sim 2\times 10^{-23} (g^\prime\zeta / 3\times 10^{-7})^4 (\sin^2 2\theta_L \cos^2 2\theta_L/0.01).$ For $\cos (2\theta_{L}) =1$, the current limit ${\rm BR}(\tau \rightarrow \mu \gamma) < 4.4 \times 10^{-8}$~\cite{pdg} implies $g' \zeta \lesssim 1.2 \times10^{-3}$, which is comparable to the neutrino trident limit~\cite{Altmannshofer:2014pba}.
                \item \textbf{LFV Higgs decay $H\to \tau \mu$:} As shown in Eqs. (\ref{11}) and (\ref{12}), we can explain the $2.4 \sigma$ hint of a nonzero Higgs width, $H\to \tau \mu$  \cite{Khachatryan:2015kon} within our model
                    $${\rm Br}(H \to \tau \mu)=\frac{{\rm Br}(H\to \tau \tau)}{2} \left( \left(\frac{\cos 2 \theta_L}{1+\sin 2\theta_L} \right)^2+ \left(\frac{\cos 2 \theta_R}{1+\sin 2\theta_R} \right)^2\right).$$
                    With $\cos 2 \theta_L=\cos 2\theta_R =0.06-0.17$,  the claimed excess ${\rm Br}(H\to \tau \mu)=(0.84^{+0.39}_{-0.37})\% $   \cite{Khachatryan:2015kon}
                    can be explained. Notice that in general $\cos^2 2 \theta_R /(1+\sin 2\theta_R)^2>0$ so
                    \be {\rm Br}(H\to \tau \mu) > \frac{{\rm Br}(H\to \tau \tau)}{2}  \left(\frac{\cos 2 \theta_L}{1+\sin 2\theta_L} \right)^2 \label{LOWER}.\ee
                    On the other hand, $\tan 2 \theta_L=(-\epsilon_{\mu \mu}+\epsilon_{\tau \tau})/(2\epsilon_{\mu \tau})$. That is we predict a lower bound on
                    ${\rm Br}(H\to \tau \mu)$ in terms of $(\epsilon_{\mu \mu}-\epsilon_{\tau \tau})/(2\epsilon_{\mu \tau})$.

            \end{itemize}

\section{ Summary and outlook\label{con}}
We have proposed a model for neutrino NSI with matter based on a new $U(1)^\prime$ gauge symmetry with a light gauge boson, $Z^\prime$.
The model by construction gives LF conserving as well as LFV terms. We have put two generations of left-handed leptons in the doublet representation of the $U(1)^\prime$ symmetry and have assumed that the third generation of the leptons are invariant under $U(1)^\prime$. In other words, in a certain basis which does not correspond to the mass basis, two generations of leptons have opposite $U(1)^\prime$ charges and the third state which can be a certain flavor (mass) eigenvector has zero $U(1)^\prime$ charge (see Eqs. (\ref{du}) and (\ref{L-opp}) for clarification). {The $U(1)^\prime$ charges of quarks are taken equal to $+1$ so we do not predict new flavor violating effects in the quark sector. The $U(1)^\prime-U(1)-U(1)$ and $U(1)^\prime-SU(2)-SU(2)$ anomalies have to be canceled by  new heavy degrees of freedom. As an example, we suggest existence of new chiral fermions with masses of order of $400-500$ GeV with the same electroweak quantum numbers as those of leptons. The existence of such new particles can be tested at accelerators such as the LHC. } {Moreover, they can provide a dark matter candidate as a bonus.}

 We have proposed two possibilities for the right-handed leptons.  In the first case, the right-handed leptons also transform as a doublet of $U(1)^\prime$ (see Eq. (\ref{R-doublet})). Thus, the SM Higgs can couple to these fields and give them mass. In the second case, the right-handed leptons transform as Eq. (\ref{R-singlet}) and two new scalar doublets are introduced to give mass to charged leptons. As a result, in the second case the branching ratios of the Higgs to charged leptons will significantly deviate from the SM prediction. Given that the measured ${\rm Br}(H \to \tau \bar{\tau})$ is in reasonable agreement with the SM prediction, the former solution is more suitable for models in which the $\tau$ lepton has nonzero $U(1)^\prime$ charge. This case is made all the more intriguing given $2.4\sigma$ hint of nonzero
${\rm Br}(H\to \tau \mu)$~\cite{Khachatryan:2015kon}, which the model can explain.
We have also shown how a seesaw mechanism for neutrino mass production can be added to this model.

{If the $Z^\prime$ coupling to the electron is nonzero,  $Z^\prime$ can decay into $e^-e^+$ which makes the detection of $Z^\prime$ in low energy luminosity frontier experiments simpler. Null results for $\phi \to \eta Z^\prime$, $Z^\prime \to e^-e^+$ and for $\pi^0\to \gamma Z^\prime$, $Z^\prime \to e^-e^+$ imply an upper bound of $\sim 10^{-3}$ on the $Z^\prime$ coupling to quarks. If the electron and electron neutrino are singlets under $U(1)^\prime$, $Z^\prime$ cannot decay into electron positron pair  so these bounds do not apply. However we have shown that the bound on ${\rm Br}(\pi^0 \to \gamma \nu \bar{\nu})$ again puts a strong bound on the $Z^\prime$ coupling to quarks for $m_{Z^\prime}<m_\pi$. However for $m_\pi <m_{Z^\prime}<2m_\mu$, the bound on the coupling of quarks to $Z^\prime$ is dramatically relaxed.}

 {In the presence of the relevant LFV gauge coupling, charged lepton $l_\beta^-$ can decay into lighter charged lepton $l_\alpha^-$ and $Z^\prime$. The longitudinal component of $Z^\prime$ leads to a strong enhancement of $(m_{l_\beta}/m_{Z^\prime})^2$ in the rate of this process which puts a strong bound
 on the LFV gauge coupling. For $m_{Z^\prime}>m_\mu$ although $\mu \to Z^\prime e$ will not be possible but the LFV gauge coupling leads to $\mu \to eee$ at tree level and again very sever bounds on the $\mu e$ component of the gauge coupling are obtained.}

 {We have estimated the maximum $\epsilon_{\alpha \beta}^{qP}$ that can be obtained within the present model still satisfying the bounds both from rare meson decays and LFV charged lepton decays. We have found $\epsilon_{\mu e}^{qP}<5\times 10^{-4}$ and $\epsilon_{\tau e}^{qP}<1.5\times 10^{-3}$. However, $\epsilon_{\mu \tau}^{qP}$ and $\epsilon_{\mu \mu}^{qP}- \epsilon_{\tau \tau}^{qP}$ for  $m_\pi <m_{Z^\prime}<2 m_\mu$ can be as large as respectively $5\times 10^{-3}$ and $0.05$ which can
be discerned by upcoming long baseline experiments.
 For this reason we mainly focus on the case that the first generation of leptons are invariant under the $U(1)^\prime$ and the second and third generation of leptons are in the doublet representation of the $U(1)^\prime$. The flavor structure of NSI in our model is shown in Eq~(\ref{fs}).}
In the parameter range in which we are interested, all the bounds on a new gauge boson  are avoided. {For example, since the coupling of $Z^\prime$ to nucleons is relatively large, the mean free path of the produced $Z^\prime$ inside supernova will be smaller than 1 cm so the $Z^\prime$ production will not directly contribute to supernova cooling.} \
 We also studied
the phenomenological effects of LFV gauge coupling on $\tau \to \mu \nu \bar{\nu}$, $\tau \to \mu +{\rm hadrons}$, $\tau \to \mu \mu \mu$ and $\tau \to \mu+\gamma$. The effects appear to be well below the sensitivity limit.

The long baseline   NO$\nu$A and DUNE experiments as well as high statistics atmospheric neutrino oscillation experiments such as IceCube Deepcore~\cite{Mocioiu:2014gua} can determine effects of neutral current NSI on neutrino oscillation pattern.
{If these neutrino experiments find that $\epsilon_{\mu \tau}$ and
$\epsilon_{\tau \tau}-\epsilon_{\mu \mu}$ are nonzero, we shall obtain a significant hint in favor of this model. The case will become stronger if the signal for $H\to \tau \mu$ is confirmed. In fact, we predict a lower bound on Br$(H\to \tau \mu)$ in terms of  $\epsilon_{\mu \tau}/(\epsilon_{\tau \tau}-\epsilon_{\mu \mu})$ (see Eq. \ref{LOWER}). {
The model can be tested by improving the bound on $\tau \to Z^\prime \mu$. Another possible test is  searching for the $Z^\prime$ production in meson decay and its subsequent decay to neutrinos (missing energy) or for $m_{Z^\prime}>m_\pi$ to $\pi \gamma$.}

\section*{Acknowledgements}
{We are grateful to the anonymous referee for pointing out the bound from $\tau \to \mu Z^\prime$.}
We are very grateful to Pedro Machado for helpful discussions.
YF would like to acknowledge partial support from the  European Union FP7 ITN INVISIBLES (Marie Curie Actions, PITN- GA-2011- 289442) { and the ICTP for the hospitality of its staff and the generous support from its associate office.  She is also grateful to INST for partial financial support. } This project has received funding from the European Union  Horizon 2020 research and innovation programme under the Marie Sklodowska-Curie grant agreement No 690575 and has received funding from the European Union  Horizon 2020 research and innovation programme under the Marie Sklodowska-Curie grant agreement No 674896. As a IGC Fellow at Penn State IMS would like to thank both the Pennsylvania State University and the Institute for Gravitation and the Cosmos for support.

\end{document}